\documentclass[prd, onecolumn, nofootinbib]{revtex4}
\usepackage{amssymb}

\usepackage{amsmath}

\usepackage{graphicx}

\usepackage{amsfonts}
\def\Ad{\displaystyle{Ad}}
\begin{document}

\title{%
Living in Curved Momentum Space }
\author{J. Kowalski-Glikman}
\email{jkowalskiglikman@ift.uni.wroc.pl}\affiliation{Institute for
Theoretical Physics, University of Wroc\l{}aw, Pl.\ Maksa Borna 9,
Pl--50-204 Wroc\l{}aw, Poland}
\date{\today}

\begin{abstract}
In this paper we review some aspects of relativistic particles'
mechanics in the case of a non-trivial geometry of momentum space.
We start with showing how the curved momentum space arises in the
theory of gravity in 2+1 dimensions coupled to particles, when
(topological) degrees of freedom of gravity are solved for. We argue
that there might exist a similar topological phase of quantum
gravity in 3+1 dimensions. Then we characterize the main properties
of the theory of interacting particles with curved momentum space
and the symmetries of the action. We discuss the spacetime picture
and the emergence of the principle of relative locality, according
to which locality of events is not absolute but becomes observer
dependent, in the controllable, relativistic way. We conclude with
the detailed review of the most studied $\kappa$-Poincar\'e
framework, which corresponds to the de Sitter momentum space.
\end{abstract}
\maketitle
\section{Introduction}

Quantum gravity is traditionally considered to be a Holy Grail of
the modern high energy physics. This theory, when constructed, would
provide a missing link between gravity and quantum, which is
necessary to complete the `unfinished
revolution'\cite{Rovelli:2006yt} of XXth century physics.
Unfortunately, the Quantum Gravity research programme faces not only
the well known tremendous technical and conceptual difficulties, but
also the seemingly complete lack of experimental feedbacks.

It has been about 15 years ago, when the quantum gravity
phenomenology programme was lounched\cite{AmelinoCamelia:1999zc}. It
was noticed that in spite of the fact that the direct observation of
Quantum Gravity effects, like scattering of elementary particles at
Planckian energies, are well beyond the reach of any foreseeable
technology, there may still exist phenomena of quantum gravity
origin that might be observable as a result of the presence of
powerful amplifiers. For example the minute effect of an interaction
of freely moving particles with the quantum space-time foam may get
amplified to an observable size if the time of flight of the
particle is long enough.

It has been rather clear from the very first days of the quantum
gravity phenomenology research programme that the most promising
class of effects to look for are those associated with possible
modifications of spacetime symmetries of special relativity. Indeed,
the flat Minkowski space (or spaces that are flat to a good
approximation at the scales of interest) is a configuration of
gravitational field with the ten parameter Poincar\'e group of
global spacetime symmetries. It is feasible that there might be
effects of quantum gravitational origin that do not vanish in the
case when the coarse grained spacetime is Minkowski. But then it is
rather natural to expect that these effects may alter the symmetries
of Minkowski spacetime, embodied in the Poincar\'e symmetry group,
somehow.

These were the motivations that served as a launching pad for two
major research projects: one positing abandoning the Lorentz
invariance, leading to construction of models with  Lorentz
Invariance Violation (LIV)
(see\cite{Mattingly:2005re,Liberati:2012th,Bluhm:2013mu} for a
review); and the second, known under the name of Doubly (or
Deformed) Special Relativity (DSR), in which much milder alternation
of the Poincar\'e symmetry was posed. In both cases it was
implicitly understood that deviations from the standard symmetries
of special relativity are of the quantum gravity origin, although
the explicit relation relations between the two was never understood
in a satisfactory way.

Doubly Special
Relativity\cite{AmelinoCamelia:2000ge,AmelinoCamelia:2000mn} was
based on the intuition that since the diffeomorphism and local
Lorentz invariances play such a central role both in both classical
and quantum gravity, the Minkowski space should still possess a
ten-parameter group of spacetime symmetries and the relativity
principle should hold even if quantum gravity effects are taken into
account. These effects (or some remnants of them) exhibit themselves
through the presence of an additional observer-independent scale of
dimension of mass (or length), which can be identified with the
Planck mass $M_{Pl}$ (or Planck length $l_{pl}$,) which {\em
deforms} the Poincar\'e symmetry algebra.  Soon after the original
papers\cite{AmelinoCamelia:2000ge,AmelinoCamelia:2000mn} appeared
some explicit models of DSR have been
proposed\cite{KowalskiGlikman:2001gp,Bruno:2001mw,Magueijo:2001cr,Magueijo:2002am}.
The reader can find more information about DSR in the
reviews.\cite{KowalskiGlikman:2004qa,KowalskiGlikman:2006vx,AmelinoCamelia:2010pd}

Although DSR was devised to be a rather general scheme, it was the
so-called $\kappa$-Poincar\'e
algebra\cite{Lukierski:1991pn,Lukierski:1992dt,Lukierski:1993wx,Majid:1994cy}
that served as a major example  and attracted a lot of attention
from the very beginning. This is a Hopf algebra with dimensionful
deformation parameter $\kappa$ of dimension of mass, which is
expected to be of order of Planck mass\footnote{Since the value of
the deformation parameter is to be derived from some fundamental
theory and/or experiments, in what follows we will use $\kappa$ to
denote the deformation parameter, whose value is not fixed, while
the term `Planck mass' will refer to $M_{Pl} =\sqrt{\hbar/G} \sim
10^{19}$ GeV.}. We will discuss the $\kappa$-Poincar\'e algebra in
more details in Sect.\ 5 below, but let us just mention its most
important feature, namely that contrary to the standard Poincar\'e
algebra of special relativity, in the case of the
$\kappa$-Poincar\'e algebra the action of both translations and
Lorentz transformation depend on momentum (and spin) of the state it
acts on. For example the wordlines of two particles with different
momenta are translated, according to $\kappa$-Poincar\'e, by a
different, momentum-dependant amounts, which means that the two
wordlines may cross for a local observer but miss each other for a
translated one.

This leads to an apparent conflict with the locality principle,
noticed first in\cite{Hossenfelder:2010tm} and further discussed
in\cite{Smolin:2010xa,AmelinoCamelia:2010qv}. In the
papers\cite{AmelinoCamelia:2011bm,AmelinoCamelia:2011pe} the lack of
absolute locality was elevated to the status of a new principle,
called {\em The principle of Relative Locality}, which states that
when quantum gravity effects are taken into account locality loses
its absolute status and becomes relative. It was the major result of
these papers to relate relative locality with curvature of momentum
space.

The idea that momentum space might be curved is quite old. It seems
that it was first spelled out in the paper\cite{born}, where it is
argued that some kind of `reciprocity principle' should be adopted,
stating that both curved spacetime and curved momentum space should
be involved simultaneously in the description of (quantum) physics.
About ten years later in the seminal paper Snyder argued that
curvature in momentum space might be necessary to handle ultraviolet
divergencies of quantum field theory\cite{Snyder:1946qz}. This paper
introduces, as a bi-product, a non-commutativity of spacetime
coordinates and the minimal length, arguing that both do not need to
be in conflict with Lorentz symmetry (for a recent review of the
minimal length scenarios see Ref.\ \cite{Hossenfelder:2012jw}.) The
ideas of Snyder was later expanded by the Russian group (see Ref.\
\cite{Kadyshevsky:1977mu} and references therein.) In the context of
DSR and $\kappa$-Poincar\'e algebra it was observed in the
papers\cite{KowalskiGlikman:2002ft,KowalskiGlikman:2003we} that both
can be naturally understood in terms of the momentum space being a
group manifold of the 4-dimensional group $AN(3)$, which as a
manifold is a submanifold of 4-dimensional de Sitter space. We will
return to this construction in Sect.\ 5.

In all these attempts the introduction of curved momentum space had
purely utilitarian character, it was aimed at solving some
outstanding problems or to provide a novel technical perspective.
The question arises however is there any fundamental reason to
believe that the momentum space is actually curved? Although the
complete story is not known one can give an argument, ultimately
relating curved momentum space with the theory of quantum gravity.

The argument is based on the intuition that the presence of a scale
is a prerequisite for emergence of a nontrivial manifold. This
intuition was beautifully expressed by Carl Friedrich Gauss already
at the dawn of differential geometry:

\begin{quote}
``The assumption that the sum of the three angles [of a triangle] is
smaller than $180^\circ$ leads to a geometry which is quite
different from our (Euclidean) geometry, but which is in itself
completely consistent. I have satisfactorily constructed this
geometry for myself [\ldots], except for the determination of one
constant, which cannot be ascertained a priori. [\ldots]  Hence I
have sometimes in jest expressed the wish that Euclidean geometry is
not true. For then we would have an absolute a priori unit of
measurement.''\footnote{As cited in \cite{Milnor}.}
\end{quote}

The necessity of the presence of the scale is easy to understand.
Indeed any nontrivial geometry requires nonlinear structures and
those can be constructed only if there is a scale that makes it
possible to construct nonlinear expressions from fundamental,
dimensionful basic variables.
 One can interpret
the Gauss' dictum as the statement that if a scale of some physical
quantity is present in a theory, one could expect that the geometry
of the corresponding manifold must be nontrivial. Or putting it in
other words: ``everything is curved unless it cannot be.''

There are several examples that support this claim. Special
relativity introduces a scale of velocity, and according to the
Gauss' dictum one suspects
 that the manifold of (three) velocities could possess
nontrivial structures. And indeed it does. Contrary to Galilean
mechanics, in special relativity the velocity composition law is
highly nontrivial
\begin{equation}\label{a}
    \vec{v}\oplus \vec{u} =
    \frac{1}{1+\vec{u}\vec{v}/c^2}\left(\vec{v}+\frac{\vec{u}}{\gamma_v}+\frac1{c^2}\frac{\gamma_v}{1+\gamma_v}\,
    (\vec{v}\vec{u})\vec{v}\right)\,, \quad
    \gamma_v=\sqrt{1-\vec{v}^2/c^2}\, .
\end{equation}
This expression is neither symmetric nor associative. It is related
to deep mathematics\cite{Girelli:2004xy} and has interesting
physical consequences (Thomas precession).

The relativistic, four-momentum space is, arguably, even more
important physically than the spacetime. Indeed virtually all
physical measurements can be reduced to the measurements of energies
and momenta of incoming particles of various kinds (probes)
performed by measuring devices located at the origin of a coordinate
system, and therefore are the momentum space, not spacetime,
measurements. It is only by observing the incoming probes that we
can infer the properties of distance events
\cite{AmelinoCamelia:2011bm,AmelinoCamelia:2011pe}. Therefore, the
momentum space measurements are physically fundamental and the
spacetime properties are inferred from them. ({\em I don't see space
\ldots I see [images of] things} the renowned Mexican painter Diego
Rivera used to say.) The question arises as to if we have good
reasons to believe that the momentum space is an almost
structureless Minkowski space, or it is conceivable perhaps that it
could posses more intricate geometrical structures?

Following the Gauss' intuition a possible way of addressing this
question is to look for a theory that could provide us with a
momentum scale $\kappa$. Such a theory indeed exists. In $2+1$
spacetime dimensions the Newton's constant $G$ has the dimension of
inverse mass raising the hope that it may provide the sought
momentum scale being a prerequisite for the emergence of a
nontrivial momentum space geometry. This hope was fully confirmed by
the dynamical model calculations and this example will be discussed
in some details in Sect.~2.

And what about gravity in the physical $3+1$ dimensions? Now the
Newton's constant is the ratio of  Planck length $l_{Pl}$ and the
Planck mass $M_{Pl}$
\begin{equation}\label{b}
    l_{Pl}=\sqrt{\hbar G}\,,\quad M_{Pl}=\sqrt{\frac\hbar G}\,,
\end{equation}
 and therefore has the dimension of length over mass.
However, one can imagine a regime of quantum gravity, in which the
Planck length is negligible, while the Planck mass remains finite.
This formally means that both $\hbar$ and $G$ go to zero, so that
both quantum and local gravitational effects become negligible,
while their ratio remains
finite\cite{Girelli:2004md,AmelinoCamelia:2011bm,AmelinoCamelia:2011pe}.
In more physical terms this regime is realized if the characteristic
length scales relevant for the processes of interest are much larger
than $l_{Pl}$, so that the spacetime quantum foamy effects can be
safely neglected, while the characteristic energies are comparable
with the Planck energy. An example of such a process might be the
gravitational scattering in the case when the longitudinal momenta
are Planckian, while the transferred momentum is very small (as
compared to $M_{Pl}$) \cite{'tHooft:1987rb}, \cite{Verlinde:1991iu}.
In the case of such processes we again encounter the situation that
the momentum scale is present, and we expect to find a nontrivial
geometry of the momentum space. Unfortunately, to date no specific
model of this kind has been formulated.

The plan of this review is as follows. In the next section we will
present a discussion of 2+1 gravity coupled to particle(s) and we
will see how curved momentum space arises in this model. We will
also briefly comment on a similar construction that might be
possible in the context of 3+1 gravity. In Sect.~3 we will present a
model o particles dynamics in the case of an arbitrary geometry of
momentum space. Sect.~4 is devoted to the discussion of spacetime
structures emerging from this construction and to relative locality.
Section 5 will present  description of $\kappa$-Poincar\'e
 momentum space geometry and the particles' model with this momentum space.

\section{Curved momentum space from gravity}

Before turning to the discussion of the most general curved momentum
space let us consider an explicit model, which will provide an
intuition as to how gravity could lead to a nontrivial geometry of
momentum space.

We start with the gravity in 2+1 dimensions. In this case the
gravitational lagrangian has the dimension of inverse length square
and thus the Newton's constant $G$ has the dimension of inverse mass
$4\pi G =\kappa^{-1}$, where $\kappa$ is the 2+1 dimensional Planck
mass. According to the argument presented in the Introduction, we
can expect that when particles become coupled to gravity their
effective momentum space will possess a nontrivial geometry, with
the characteristic scale $\kappa$. Let us see explicitly how this
comes about.

It has been shown by Witten\cite{Achucarro:1987vz,Witten:1988hc}
that gravity in 2+1 spacetime dimensions can be formulated as a
Chern-Simons topological field theory with the gauge group being (in
the case of vanishing cosmological constant) the 2+1 dimensional
Poincar\'e group $ISO(2,1)$, whose generators satisfy the following
commutational relations
$$
[J_a, J_b] = \epsilon_{abc}\, J^c\,,\quad [J_a, T_b] =
\epsilon_{abc}\, T^c\,,\quad [T_a, T_b] =0\,.
$$
In what follows we will call $J_a$ and $T_a$ Lorentz and
translational generators, respectively. In terms of these generators
the connection one-form decomposes into spin connection $\omega^a$
and dreibein $e^a$
\begin{equation}\label{2.1}
A = \omega^a\, J_a + e^a \, T_a\,
\end{equation}
and the action takes the form
\begin{equation}\label{2.2}
    S = \frac{1}{4\pi G} \int \left< A\wedge dA + \frac23\, A\wedge A \wedge
    A\right>\,,
\end{equation}
where $\left<\star\right>$ denotes an invariant inner product on the
Poincar\'e algebra defined by\cite{Witten:1988hc}
\begin{equation}\label{2.3}
    \left<J_a T_b\right> = \eta_{ab}\,,\quad \left<T_a T_b\right> = \left<J_a J_b\right>
    =0\,.
\end{equation}
One can show by direct calculation\cite{Witten:1988hc} that this
action with the connection given by (\ref{2.1}) reduces to the
standard action for gravity in 2+1 dimensions.

It will be convenient to what follows to  assume that the 2+1
manifold ${\cal M}$ has the product structure of time times space
${\cal M} = \mathbb{R}\times {\cal S}$ and to decompose the
connection $A$ accordingly
\begin{equation}\label{2.4}
    A = A_0\, dt + A_s\,,\quad A_s = A_i\, dx^i\,, \quad i=1,2
\end{equation}

Let us now turn to the coupling of 2+1 gravity to a point
particle\cite{Witten:1988hf}. One can use fix the diffeomorphism
invariance on the particle's wordline in such a way so as to put the
particle into rest at the origin $\mathbf{x}=\mathbf{0}$. In this
case the particle is characterized by its mass $m$ and spin $s$,
both being  the charges associated with the group of spacetime
symmetries and therefore can be written collectively as an element
of the Cartan subalgebra of the gauge algebra $Q= mJ_0 + sP_0$. In
the following, for simplicity, we will assume that the particle's
spin vanishes, $s=0$.

The particle at rest is fully characterized by its wordline and the
Lie algebra valued charge $Q$, and the simplest, minimal coupling to
gravity takes the form
\begin{equation}\label{2.5}
    S^{(0)}_{int} = \int d^3x \left<A_0\, Q\right>\delta^2(\mathbf{x})
\end{equation}
The action (\ref{2.5}) is not only manifestly diffeomorphism
not-invariant, it also breaks the gauge symmetry on the particle's
wordline. This is actually a desired feature of the formalism,
because, as we will see, the gauge degrees of freedom along the
wordline become the dynamical degrees of freedom of the particle.
Since the  away from the particle the theory is topological and
degrees of freedom of gravity  are pure gauge, the degrees of
freedom along the wordline are the only `real'  degrees of freedom
in the theory (if the space manifold $\cal S$ has simple enough
topology.)

 Indeed plugging the gauge transformed field
\begin{equation}\label{2.6}
    A_0^\mathfrak{h} = \mathfrak{h}^{-1}\, A_0 \mathfrak{h} + \mathfrak{h}^{-1}\, \dot{
    \mathfrak{h}}
\end{equation}
we get
\begin{equation}\label{2.7}
    S_{int} = \int d^3x \left<A_0\, \mathfrak{h}\, Q \,
    \mathfrak{h}^{-1}\right>\delta^2(\mathbf{x}) + \int dt \left<\mathfrak{h}^{-1}\,
    \dot{
    \mathfrak{h}}\, Q\right>\,.
\end{equation}
The first term describes the coupling of the particle with the
gravitational field, while the second is the particle's kinetic
term. To see this let us decompose the Poincar\'e group element
$\mathfrak{h}$ into Lorentz and translational parts
$\mathfrak{h}=(\mathfrak{u}, \mathfrak{x})$, $\mathfrak{q}=q^a\,
T_a$in terms od which the second term in (\ref{2.7}) takes the form
\begin{equation}\label{2.8}
    \int dt\, m \left<\dot{ {q}}^a\, T_a\,
    \mathfrak{u}\, J_0 \mathfrak{u}^{-1}\right>= \int dt\, \dot q^a\, p_a\,.
\end{equation}
where we define the momenta $p_a$ by the formula $m\,\mathfrak{u}\,
J_0 \mathfrak{u}^{-1} \equiv p_a\, J^a$. This procedure has clear
physical interpretation. A Poincar\'e group element describe Lorentz
transformation and translation. We use the Lorentz part to boost the
particle from its rest state to the actual state of motion
characterized by momentum $p_a$; the translation moves the
particle's position from the origin to its actual position $q^a$.

As mentioned already gravity in 2+1 dimensions is described by a
topological field theory, and therefore it does not posses local
degrees of freedom: neither Newtonian attraction nor gravitational
waves are possible. This clearly follows from the fact that the
field equations following from (\ref{2.2}) force both the curvature
of $\omega$ and torsion to vanish and therefore the spacetime of 2+1
gravity, locally at least, is flat. In the case of the particles
coupling the situation is similar: the field equations says that the
curvature and torsion are zero everywhere except at the very
position of the particle, where they acquire a delta-like
singularity\cite{Staruszkiewicz:1963zza}. This does not mean however
that all the dynamics is gone. The particle's dynamical degrees of
freedom are still there. The presence of gravity does not change the
number of degrees of freedom in the theory. It does however a
remarkable thing: gravitational degrees of freedom are absorbed by
the particle's ones effectively deforming its the kinematics and
dynamics.

Let us sketch how this comes about (we refer the reader to the
upcoming paper\cite{JKGTT} for more details and detailed discussion
of the subtle points of the construction). We start with the kinetic
term of the action (\ref{2.2}) written down with the help of the
decomposition (\ref{2.4})
\begin{equation}\label{2.9}
    S_{kin} = \frac1{4\pi G}\int dt\, \int_{\cal S}\left< A_s \wedge \dot
    A_s\right>+   \left< m\, J_0, \mathfrak{h}^{-1}\dot{
   \mathfrak{h}}\right> \, \delta^{(2)}(\mathbf{x})\,,
\end{equation}
where the connection one-form $A_s$ is constrained by the Gauss law
(the field equation of $A_0$)
\begin{equation}\label{2.10}
    \frac1{2\pi G} F(A_s) = =  m\, \mathfrak{h}\,  J_0\, \mathfrak{h}^{-1}\, \delta^{(2)}(\mathbf{x})dx^1\wedge
    dx^2\,.
\end{equation}

The constraint (\ref{2.10}) can be easily solved as follows. Let us
decompose the 2-dimensional space manifold $\cal S$ into two
subregions: the plaquette $\cal D$ being a circle with the center at
the position of the particle, on which we introduce coordinates
$0\geq r\geq 1$ and $0\geq\phi\geq 2\pi$ and the asymptotic region
${\Sigma}$ with $r\geq 1$. These two regions have a common boundary
$\cal H$, $r=1$, $0\geq\phi\geq 2\pi$. On the asymptotic region the
connection is flat and the gauge field takes the form
\begin{equation}\label{2.11}
    A_{\Sigma} = \mathfrak{g}^{-1}\, d\mathfrak{g}
\end{equation}
where $\mathfrak{g}$ is an element of the Poincar\'e gauge group.
One can find a general solution of (\ref{2.10}) on the disc as well,
it reads\footnote{This follows immediately from the fact that
$F(A^\mathfrak{g})=\mathfrak{g}^{-1} F(A) \mathfrak{g}$, where
$A^\mathfrak{g}$ denotes the gauge transformed connection and
$F(Gm\, J_0 d\phi) = Gm\, J_0 dd\phi= 2\pi Gm J_0 \delta(\mathbf{x})
dx^1\wedge dx^2$.}
\begin{equation}\label{2.12}
    A_{\cal D} =G\, m\,\bar{\mathfrak{g}}^{-1}\, J_0\bar{\mathfrak{g}} d\phi+\bar{\mathfrak{g}}^{-1}\,
    d\bar{\mathfrak{g}}\,, \quad \mathfrak{g}(0)=\mathfrak{h}^{-1}\,.
\end{equation}
The fact that (\ref{2.12}) is a general solution of (\ref{2.10}) on
the disk can be easily checked using the identity $dd\phi = 2\pi
\delta(\mathbf{x}) dx^1\wedge dx^2$. In addition we assume that the
gauge field is continuous across the boundary $\cal H$, which
imposes an additional relation between $\mathfrak{g}$ and
$\bar{\mathfrak{g}}$
\begin{equation}\label{2.13}
    \left. \mathfrak{g}^{-1}\, d\mathfrak{g}\right|_{\cal H}= \left. G\, m\,\bar{\mathfrak{g}}^{-1}\, J_0\bar{\mathfrak{g}} d\phi+\bar{\mathfrak{g}}^{-1}\,
    d\bar{\mathfrak{g}}\right|_{\cal H}\,.
\end{equation}
Decomposing ${\mathfrak{g}} = (\mathfrak{u}, \mathfrak{q})$,
$\bar{\mathfrak{g}} = (\bar{\mathfrak{u}}, \bar{\mathfrak{q}})$ as
before, a general solution of (\ref{2.13}) can be found to
read\cite{Meusburger:2005mg}
\begin{equation}\label{2.14}
\mathfrak{u}= \mathfrak{n}\, e^{Gm  J_0\phi}\,\bar{
\mathfrak{u}}\,,\quad \mathfrak{q}= \eta + \Ad(\mathfrak{n}\, e^{Gm
J_0\phi})\,\bar{\mathfrak{q}}\,.
\end{equation}
where $\mathfrak{n}$ and $\eta$ are time-dependent (but
space-independent) Lorentz and translation groups elements,
respectively. Notice that in the transition from (\ref{2.12}) to
(\ref{2.13}) the Lie algebra valued momentum of a free particle at
rest $m\, J_0$ (cf.\ (\ref{2.6}) and discussion that follows)
becomes replaced by group valued one $e^{Gm J_0\phi}$, when
gravitational gauge degrees of freedom are taken into account.

A remarkable thing happens when the explicit forms of the
Chern--Simons connection (\ref{2.11}), (\ref{2.12}) with the
boundary conditions (\ref{2.14}) are substituted to the action
(\ref{2.9}). Namely, the action of the gravity plus particle system
collapses to the action describing a deformed particle with curved
momentum space being an $SO(2,1)$ group manifold. Explicitly, in
terms of the group valued momenta
\begin{equation}\label{2.15}
    \Pi\equiv \mathfrak{u}\, e^{2\pi\, Gm  J_0}\, \mathfrak{u}^{-1} = p_3\, \mathbf{1}
    + \frac{p_a}\kappa \, J^a\,,
\end{equation}
where the momenta are coordinates on the group manifold of SO(2,1),
which geometrically is a 2+1 dimensional anti de Sitter space
\begin{equation}\label{2.16}
    p_3^2 - p_ap^a = \kappa^2\,,\quad \kappa = (2\pi G)^{-1}
\end{equation}
subject to the mass shell condition\footnote {In this paper we use
the signature $(-,+,+)$ in 2+1 dimensions and $(-,+,+,+)$ in 3+1
dimensions.}
\begin{equation}\label{2.17}
     p_ap^a = -m^2\,.
\end{equation}
This effective kinetic term reads
\begin{equation}\label{2.18}
    S_{kin} = \int dt \left<\Pi^{-1} \, \dot\Pi T_a\right> x^a +
    N(p^2+m^2)\,,
\end{equation}
where is a standard Lagrange multiplier enforcing the mass shell
constraint. Notice that the kinetic term is written in terms of the
Kirillov symplectic form\cite{Kirillov}, which is a natural
symplectic form on a group manifold. In components the action
(\ref{2.18}) this effective kinetic term reads
\begin{equation}\label{2.19}
    S_{kin} = -\int dt p_3 \dot p_a x^a + \frac{1}\kappa\epsilon^{abc}\, \dot
    p_a\,    x_b\,p_c-\frac{1}{\kappa^2}\, \dot p_a\, p^a p_b\, x^b+
    N(p^2+m^2)\,,\quad p_3^2 = 1 + \frac{1}{\kappa^2}\, p_ap^a
\end{equation}
Notice that in the limit $\kappa\rightarrow\infty$ (i.e., in the
no-gravity limit $G\rightarrow0$) this action reproduces the free
relativistic particle action (\ref{2.8}).

Anticipating  the discussion of a more general case to be presented
in the next section, let us notice that the action (\ref{2.19}) can
be neatly written down as
\begin{equation}\label{2.20}
    S_{kin} = -\int dt  \dot p_\alpha\, E^\alpha{}_b(p)\, x^a +
    N(p^2+m^2)\,,
\end{equation}
where $E^\alpha{}_b(p)$ is a dreibein on the momentum space, and to
stress that the momentum space is curved we used the Greek index
$\alpha$ to label components of the momentum.

It is not completely obvious if the construction presented above can
be extended without modifications to the case of many particles, but
this happen to be true nevertheless\cite{Meusburger:2005mg}: for the
finite number of particles the kinetic term turns out to be a sum of
terms of the form (\ref{2.18}). This construction can be extended
further to the case of a (not self interacting) scalar field coupled
to gravity\cite{Freidel:2005bb,Freidel:2005me}.

Till now we discussed only free particle(s). Let us now consider a
bunch of particles, which, in addition to their coupling to the
gravitational field considered above have some other interactions of
not topological nature. Assume that these interactions are contact
in a sense that there is a well localized vertex in which the
interaction takes place and one can think of asymptotic regions
where the interactions can be neglected. Let us consider the
simplest possible, nontrivial vertex in which two incoming particles
interact and form a single outgoing particle. The law of momentum
conservation at the vertex is the postulate that the total momentum
of the initial two particle configuration equals the momentum of the
outgoing one. Therefore the problem of finding the momentum
conservation rule reduces to the problem to find out what is the
correct measure of total momentum of the two-particles initial
state. Since, as we have seen, the momenta become coordinates on a
group manifold it is natural to expect that the total momentum is
given by the product of the corresponding group
elements\cite{Bais:2002ye}
\begin{equation}\label{2.21}
    \exp\left(p_a^{(tot)}\, J^a\right) = \exp\left(p_a^{(1)}\,
    J^a\right)\,\exp\left(p_a^{(2)}\,
    J^a\right)\,,
\end{equation}
which can be generalized to the case of an arbitrary group momentum
space. Introducing the operation $\oplus$ such that
\begin{equation}\label{2.22}
    p_a^{(tot)}=p_a^{(1)}\oplus p_a^{(2)}
\end{equation}
we define it by demanding that
\begin{equation}\label{2.23}
    g\left( p^{(1)}\oplus p^{(2)}\right)=g\left( p^{(1)}\right)g\left(
    p^{(2)}\right)\,.
\end{equation}

The case of 3+1 gravity is by far less understood but there are some
partial results that make it reasonable to claim that there may
exist a regime of this theory quite similar to the 2+1 dimensional
case discussed above.

In 3+1 dimensions gravity is certainly {\em not} described by a
topological field theory, as it was the case in 2+1 dimensions, but
it turns out that it is surprisingly `close' to such a theory. In
fact, the 3+1 gravity can be described by a `constrained' BF theory,
which means that Einstein lagrangian can be written down as a sum of
the topological BF theory lagrangian and a small symmetry breaking
term (for a recent review and extensive references to earlier works
see Ref.\ \cite{Freidel:2012np}.)

Specifically, one starts the construction introducing the connection
of de Sitter gauge group
\begin{equation}\label{2.24}
    A = \omega^{ij}\, T_{ij} + \frac1\ell\, e^i\, T_{i4}\,,
\end{equation}
where $T_{IJ} = (T_{ij}, T_{i4})$, $I,J=0,\ldots,4$,
$i,j=0,\ldots,3$ are dimensionless generators of the $SO(4,1)$ (de
Sitter) group and $\ell$ is a length scale needed for the
dimensional reasons, because we want the tetrad to be
dimensionless\footnote{The canonical dimension of the gauge field
potential is inverse length, thus the connection one form is
dimensionless.}. We also introduce a $so(4,1)$ Lie algebra valued
two form field $ B =B^{AB}\, T_{AB}$. In terms of these fields the
action reads\cite{Smolin:1998qp,
 Freidel:2005ak}
 \begin{equation}\label{2.25}
    S =\frac1{16\pi}\, \int_{\cal M} B^{IJ} \wedge F_{IJ}(A) -
\frac{\beta}{2}\,B^{IJ} \wedge B_{IJ}- \frac{\alpha}4\,
\epsilon_{ijkl}\,
        B^{ij}\wedge B^{kl}
  \,.
 \end{equation}
The first two terms (for any value of the parameter $\beta$) form
the action of the topological BF theory. The last term manifestly
breaks the gauge symmetry down to the Lorentz group $SO(3,1)$ and is
essential for the emergence of local degrees of freedom of gravity.
Because of the presence of this last term the theory described by
(\ref{2.25}) is called the `constrained BF theory.' It is worth
noticing that when the limit $\alpha\rightarrow0$ is taken
(\ref{2.25}) to the action of a pure topological theory with no
dynamical degrees of freedom.

It comes as a pleasant surprise that the theory described by
(\ref{2.25}) is in fact equivalent to the Einstein-Cartan theory. To
see this one has to solve the algebraic equations for $B$ field and
substitute the result back to the action. As a result one gets the
Holst action\cite{Holst:1995pc}
\begin{equation}\label{2.26}
    32\pi G\, S=\int R^{ij}\wedge e^{k}\wedge e^{l}\,\epsilon_{ijkl}+\frac{\Lambda}{6}\int  e^{i}\wedge e^{j}\wedge e^{k}\wedge e^{l}\,
    \epsilon_{ijkl} +\frac{2}{\gamma}\int R^{ij}\wedge e_{i}\wedge
    e_{j}\,.
\end{equation}
supplemented by topological invariants: Euler, Pontryagin, and
Nieh-Yan classes\cite{Freidel:2005ak}. The physical coupling
constants, the Newton's constant $G$ and the Immirzi-Barbero
parameter $\gamma$ are related to the coupling constants of the
original action (\ref{2.25}) as follows
\begin{equation}\label{2.27}
    G = \frac{\alpha^2 + \beta^2}\alpha\, \frac1\Lambda\,,\quad
    \gamma=\frac\beta\alpha\,,\quad \Lambda = \frac3{\ell^2}
\end{equation}
It can be shown that the constrained BF theory described by
(\ref{2.25}) can be coupled to point particles with momentum and
spin exactly as it was done for 2+1 gravity
above\cite{Freidel:2006hv}.

In the introductory section we argued that the curved momentum space
may arise if one takes the limit of quantum gravity
$\ell_{Pl}\rightarrow0$ with $M_{Pl}$ kept fixed, or putting it in
another way, $G\rightarrow0$, $\hbar\rightarrow0$ with their ratio
fixed. Having the theory (\ref{2.25}) coupled to particles one can
look for curved momentum space in the effective theory arising in
the topological limit $\alpha\rightarrow0$, which reminds in many
respects the 2+1 dimensional situation. Notice that according to
(\ref{2.27}) this is again the limit of zero gravitational and
Planck constants, so it is possible that the theory operating in
this limit may describe a curved momentum space, at least in some
circumstances. Some more quantitative works based on this general
intuition have been done in the
past\cite{KowalskiGlikman:2006mu,KowalskiGlikman:2008fj}, but the
results were not conclusive.

In the rest of this review we will assume that there are physical
systems that require the curved momentum space for their description
and we will present the techniques necessary to analyze properties
of such systems.

\section{Relativistic particles with curved momentum space}

In the previous section we saw how in 2+1 spacetime dimensions the
free relativistic particle action becomes deformed by absorbing
topological degrees of freedom of gravity. The net result of this
absorbtion process is that the momentum space of the particle
becomes a curved manifold (a group manifold) and the components of
the particle's momentum become (coordinate) functions on this
manifold. In the case of the physical 3+1 spacetime dimensions the
analogous derivation is still missing (although, as we saw, there
are some arguments that it may work as well) and therefore here we
will assume that the geometry of momentum space is arbitrary. This
will make it possible to identify the geometric object that are
necessary to make the transition from the standard  formulation of
the theory of relativistic particles to the one appropriate to the
case of a nontrivial momentum space geometry.

To see how this generalization can be implemented let us start with
the discussion of the action of a free relativistic particle. It
reads
\begin{equation}\label{3.1}
    S^0=-\int_{-\infty}^{+\infty} d\tau\, x^a\dot p_a
    +N\left(\eta^{ab}p_ap_b+m^2\right)\,,
\end{equation}
where the overdot denotes differentiation with respect to the
parameter $\tau$. The lagrangian in (\ref{3.1}) consists of two
terms: the kinetic one $-x^a\dot p_a$, $a,b=0,\ldots,3$  and the
mass shell constraint $\eta^{ab}p_ap_b+m^2$ imposed by the Lagrange
multiplier $N(\tau)$. It will be important for the later purposes to
note that the term $\eta^{ab}p_ap_b$ is nothing but the square of
the Minkowski distance between the point ${\cal P}$ in momentum
space, with coordinates $p_a$ and the momentum space origin ${\cal
O}$ with coordinates $p_a=0$, calculated along the straight line
joining these two points, i.e., along the geodesic of the Minkowski
space geometry.

Let us note in passing that the action (\ref{3.1}) is manifestly
invariant under global Lorentz and local $\tau$ reparametrization
symmetries, as well as  global translations
\begin{equation}\label{3.2}
   \delta_\xi x^a = \xi^a\,,\quad \delta_\xi p_a=\delta_\xi N =0\,.
\end{equation}
The equations of motion resulting from  (\ref{3.1}) are
\begin{equation}\label{3.3}
    \dot p_a = 0\,,  \quad \dot x^a = 2N\,
    \eta^{ab}p_b\,,\quad \eta^{ab}p_ap_b=-m^2\,.
\end{equation}
The first equation is the momentum conservation, the second relates
velocity to momentum, while the third is the mass-shell condition.

The system of several non-interacting relativistic particles labeled
by $\cal I$ is described by the action being a sum of the actions
(\ref{3.1}),
\begin{equation}\label{3.4}
S^0_{free}=-\sum_{{\cal I}}\int d\tau \,x_{\cal I}^a\dot p^{\cal
I}_a
    +N_{\cal I}\left(\eta^{ab}p^{\cal I}_ap^{\cal I}_b+m_{\cal
    I}^2\right)\,.
\end{equation}

We can introduce particles interactions as follows. Let some number
of wordlines meet at the  interaction vertex, and let us assume that
at the vertex the momenta are conserved. One can further assume that
for the wordlines corresponding to incoming particles the parameter
$\tau$ has the range from $-\infty$ to $0$, while for the outgoing
ones from $0$ to $\infty$, and the interaction point corresponds to
the $\tau=0$ on each wordline, but from now on we will not write the
range explicitly.

To include the interaction one adds to the action (\ref{3.4}) an
interaction term of the form\cite{AmelinoCamelia:2011bm}
\begin{equation}\label{3.5}
S^0_{int}=    z^a\,{\cal K}_a(p^{\cal I})\,,\quad {\cal K}_a(p^{\cal
I})\equiv\widetilde{\sum_{{\cal I}}}p^{\cal
    I}_a
\end{equation}
where the tilde over sum indicates that the incoming momenta are
taken with plus, while outgoing with minus signs. We assume that the
total action is $S_{tot}=S_{free}+S_{int}$. Along the wordlines the
equations of motion following from this total action do not change
and take the form (\ref{3.3}) for each particle and variation over
$z^a$ results in the momentum conservation rule at the vertex.
However, since the wordlines are semi-infinite now, when varying the
free action over momentum, from each particle action we get a
boundary term that gets combined with the variation of ${\cal K}_a$
leading to the condition
\begin{equation}\label{3.6}
    x_{\cal
I}^a(0) = z^a\quad \forall {\cal I}\,.
\end{equation}
This equation says that the `local interaction coordinate' $z^a$ is
equal to the coordinates of the ends of the wordlines $x_{\cal
I}^a(0)$. It is worth noticing that eq.\ (\ref{3.6}) is covariant
under Lorentz transformations and invariant under translations, if
we take $\delta_\xi z^a = \xi^a$. We see therefore that in special
relativity locality is {\em absolute}: if an event (interaction of
particles) is local for one inertial observer it is local for other
inertial observers, in the sense that for all observers $x_{\cal
I}^a(0)= x_{\cal J}^a(0)= z^a$ for all ${\cal I}, {\cal J}$.

It should be noted that the absolute locality exhibited by the
relativistic particles model relies on the `correct' choice of the
coordinates on the particles' phase space. Indeed, had we chosen
another position coordinates $x^a \mapsto X^a \equiv M^a_b(p) \,
x^b$ the theory would suffer from an apparent lack of locality:
under translation the wordlines would transform in a momentum
dependent way, i.e., instead of (\ref{3.2}) we would have $\delta
X^a =M^a_b(p) \, \xi^b$. We will return to this point in Sect.\ 4
below while discussing relative locality.

Now we want to generalize this setup to the case of curved momentum
space\cite{AmelinoCamelia:2011bm}. As we saw in the previous section
to define the kinetic term one needs to employ the momentum space
tetrad $E^\alpha_a(p)$\footnote{From now on, to stress that momenta
are coordinates on a curved manifold we label them with a Greek
index.}. As for the dispersion relation we noticed above that in the
case of special relativity it can be interpreted as a geodesic
distance in Minkowski space between the spacial `zero momentum'
point $\cal O$ with coordinates $p_\alpha=0$ and the point $\cal P$
with the actual coordinates $p_\alpha$. In the case of a curved
momentum space manifold this generalizes to a square of the geodesic
distance ${\cal C}=D^2(p)$, calculated with the help of the metric
$g^{\alpha\beta}(p)=E^\alpha_a(p)\, E^\beta_b(p)\, \eta^{ab}$. Taken
all this together we get
\begin{equation}\label{3.7}
S^\kappa_{free}=-\sum_{{\cal I}}\int d\tau \,x_{\cal I}^a\,
E^\alpha_a(p^{\cal I})\dot p^{\cal I}_\alpha
    +N_{\cal I}\left({\cal C}(p^{\cal I})+m_{\cal
    I}^2\right)\,.
\end{equation}
Assuming invertibility of the momentum space tetrad
$E^\alpha_a(p^{\cal I})$ the equations of motion for $x_{\cal I}^a$
force the momenta to be constant along the wordlines, $\dot p^{\cal
I}_\alpha=0$. Further from the equation for $N$, it follows that the
deformed mass shell condition ${\cal C}(p)=-m^2$ has to be
satisfied. Finally, the equation for $p^{\cal I}_\alpha$  imposes
the following relation between momenta and velocities
\begin{equation}\label{3.8}
    {\dot x}_{\cal I}^a = N\, E_\alpha^a(p^{\cal I})\,\frac{\partial}{\partial p^{\cal I}_\alpha}\,{\cal C}(p^{\cal
    I})\,.
\end{equation}
Notice that $N$ can be reabsorbed into redefinition of the parameter
$\tau\rightarrow \bar\tau$, $d\bar\tau/d\tau=N$, as usual, so it can
be gauge fixed to be equal to an arbitrary positive number.

One can calculate the Poisson brackets on the phase space of the
theory described by (\ref{3.7}) by simply noticing that the pairs
$(x_{\cal I}^\alpha(p), p^{\cal I}_\beta)$, where $x_{\cal
I}^\alpha(p)= x_{\cal I}^a\, E^\alpha_a(p^{\cal I})$ have the
canonical Poisson bracket $$\{x^\alpha_{\cal I},p^{\cal
J}_\beta\}=\delta_{\cal I}^{\cal J}\, \delta^\alpha_\beta\,,$$ from
which the brackets for $(x_{\cal I}^a, p^{\cal I}_\beta)$ can be
read off (see (\ref{4.1}) and (\ref{4.2}) below for explicit
expressions.)
\newline

Having discussed the deformed free particles' action let us now turn
to interactions. Recalling eq.\ (\ref{3.5}) we see that in order to
define the momentum conservation rule at the vertex we must first
introduce the operation $\oplus$ such that the total momentum of two
particles having momenta $p_\alpha$ and $q_\alpha$ is
\begin{equation}\label{3.9}
    p^{tot}_\alpha=(p\oplus q)_\alpha\,.
\end{equation}
In principle the operation $\oplus$ does not need to be neither
symmetric (i.e., $p\oplus q$ could be not equal to  $q\oplus p$~)
nor associative (i.e.,  $(p\oplus q)\oplus r$ could be not equal to
$p\oplus (q\oplus r)$. ) In the case of the momentum space being a
group manifold, $\oplus$ is constructed from the group composition
rule and is associative (because the group composition is) but not
symmetric (for non-abelian groups.) We also need the operation of
inverse (or antipode in the language of Hopf algebras) denoted
$\ominus$ which generalizes the `minus' and satisfies
\begin{equation}\label{3.10}
    p\oplus(\ominus p) = (\ominus p)\oplus p=0
\end{equation}
Using the operators $\oplus$ and $\ominus$ we can generalize the
form of the interaction term, eq.\ (\ref{3.5}) to read
\begin{equation}\label{3.11}
S^\kappa_{int}=    z^\alpha\,{\cal K}_\alpha(p^{\cal I})\,,\quad
{\cal K}_\alpha(p^{\cal I})\equiv\left(\widetilde{\bigoplus_{{\cal
I}}}p^{\cal
    I}\right)_\alpha
\end{equation}
and the analogue of eq.\ (\ref{3.6}) takes the form
\begin{equation}\label{3.12}
    x_{\cal I}^a(0) = E^a_\beta\, z^\alpha\, \frac{\partial{\cal
    K}_\alpha}{\partial p_\beta^{\cal I}}\,.
\end{equation}
Let us comment at this point that to define ${\cal K}$ we use the
operator $\tilde\bigoplus$ with outgoing particles  contributing
with the $\ominus$ to the sum. This corresponds to the physical
assumption that the velocities of the particles are timelike, future
directed for both incoming and outgoing particles. Alternatively one
can use the $\bigoplus$ operation, with the understanding that for
the outgoing particles the velocity is past directed.

Let us finish this part with some  comments on the theory of
interacting particles with curved momentum space formulated above.

\begin{itemize}

\item It should be stressed once again that he construction
presented above would not be possible had it not for the presence of
the momentum scale. This is clearly visible if one attempts to
expand the functions introduced above in powers of moment, near the
origin $\cal O$, $p=0$. For momentum space tetrad we have, for
example
$$
E_a{}^\alpha(p) = \delta_a^\alpha + \frac{1}{\kappa}
C^{\alpha\beta}{}_a\, p_\beta + \ldots
$$
and since $E_a{}^\alpha(p)$ is dimensionless by definition $\kappa$
must have dimension of mass.

\item The momentum conservation rule (\ref{3.11}) is, in general
neither symmetric, nor associative nonlinear function of the momenta
$p^{\cal I}$. This means that  for  given momenta of the incoming
particles there might be many reaction channels with outgoing
momenta being different in each. However this effect will be at
least of order $1/\kappa$ and is certainly not in conflict with any
available experimental data. Moreover, the existence of many
channels of momentum conservation does not contradict any
fundamental physical postulate.

\item As discussed in details in \cite{AmelinoCamelia:2011bm} and
\cite{Freidel:2011mt} there are interesting and deep geometrical
structures behind the momentum composition rule. As shown there any
non-trivial momentum composition can be geometrically expressed in
terms of a connection on momentum manifold and vice versa, any
connection defines a composition rule. This connection is defined
(at the origin ${\cal O}$; the expression at arbitrary point can be
find in\cite{AmelinoCamelia:2011bm,Freidel:2011mt})
$$
\Gamma^{\alpha\beta}_\gamma= -\left.\frac{\partial^2}{\partial
p_\alpha\partial q_\beta}\, (p\oplus q)_\alpha\right|_{p=q=0}
$$
If the composition rule is not symmetric, the connection has torsion
$$
T^{\alpha\beta}_\gamma=\Gamma^{\alpha\beta}_\gamma-\Gamma^{\beta\alpha}_\gamma\,;
$$
if it is not associative, it has a non-vanishing curvature
$$
R^{\alpha\beta\sigma}_\gamma =\left.\frac{\partial^3}{\partial
p_{[\alpha}\partial q_{\beta]}\partial r_\gamma}\, \left[((p\oplus
q)\oplus r)_\alpha- (p\oplus (q\oplus
r))_\alpha\right]\right|_{p=q=r=0}
$$
Moreover, the metric introduced in the course of defining the free
action (\ref{3.7}) does not need to be (and usually is not)
connection compatible, i.e., the covariant derivative of the
connection applied to the metric does not vanish in general
$\nabla^\alpha g^{\beta\gamma}\equiv N^{\alpha\beta\gamma}\neq0$. It
turns out that the abstract theory of momentum is closely related to
the mathematical theory of loops, see \cite{Girelli:2010zw} and
\cite{Freidelnotes} for more details and references to the original
mathematical literature.

It is worth noticing that in the best understood, and physically
most important cases, the momentum composition law is a group
product, which is, by definition associative and, in general,
non-commutative. It follows that in the case of group-like momentum
composition law the curvature vanishes, but the torsion is generally
not zero. Usually the metric one uses to define the free particle
action is not connection-compatible.

\item It should be stressed once again that it follows from eq.\
(\ref{3.12}) that contrary to the special relativistic case
(\ref{3.5}) the value of $x^a$ at the end of the trajectory depends
on the momenta of all particles that interact in the vertex
(including its own, in general). This is the signal of a completely
new spacetime physics and  emergence of relative locality, which we
will discuss below.

\item Last but not least let us comment on an important conceptual
problems that apparently plagues any theory with a nonlinear
momentum space deformation. One could argue that such deformations
clearly contradict the everyday experience that macroscopic bodies
satisfy the standard linear on-shell relations and linear momentum
conservation rules. And yet the momenta of macroscopic bodies are by
many orders of magnitude larger than the scale $\kappa$, and thus
for them the nonlinear deformation terms should be particularly
large. This problem is dubbed {\it the soccer ball problem}. To
resolve this paradox one notices that macroscopic bodies are
composite systems, built from large number $N$ of elementary
constituents. Then it can be argued that, for example, in the
macroscopic scattering problem we have to do not with one
interaction process, but with a huge number of elementary
interactions, between elementary constituents, which as a result
re-scale the deformation scale to become of order of $N\kappa$. Thus
for a soccer ball the deformation scale is at least of order of
$10^{23} M_{Pl} \sim 10^{18}$g (see \cite{AmelinoCamelia:2011uk} for
more details and references.) Let us notice at this point that as
another side of the same coin we can ask ourselves if we have good
reason to believe that the Standard Model elementary particles are
indeed elementary or `soccer bolls' of some kind, or, in another
words, do we have good reason to believe that the scale $\kappa$ is
indeed of order of the Planck mass. The answer to this question
rests with the experiment.

\end{itemize}

\subsection{Symmetries of the action}

As we briefly discussed above, the action of a particle in special
relativity is invariant under global translational and Lorentz
symmetries. Let us now investigate what is the fate of these
symmetries in the theory of particles with curved momentum space.

Consider first the translational symmetry. For simplicity let us
assume that we have 3 particles meeting at the vertex labeled by the
interaction coordinate $z^\alpha$ and that the range   of the
parameter $\tau$ is $(-\infty,0)$ for all three wordlines. As in the
case of special relativity the translational symmetry leaves the
momenta invariant $\delta_\xi p^{\cal I} = \delta_\xi N^{\cal I}=0$.
Therefore the condition for translational invariance takes the form
$$
\sum_{{\cal I}}\int_{-\infty}^0 d\tau \,\delta x_{\cal I}^a\,
E^\alpha_a(p^{\cal I})\dot p^{\cal I}_\alpha - \delta
z^\alpha\,{\cal K}_\alpha(p^{\cal I})=0\,.
$$
If we now take
\begin{equation}\label{3.13}
    \delta x^a_{\cal I}=\xi^\gamma\, \frac{\partial {\cal
    K}_\gamma}{\partial p_\beta^{\cal I}}\, E^a_\beta
\end{equation}
the sum of integrands over ${\cal I}$  gives a total derivative $d
(\xi^\gamma{\cal K}_\gamma)/d\tau$ and the translational invariance
condition reduces to $\delta z^\alpha = \xi^\alpha$. The wordline
labels $x^a_{\cal I}$ transform therefore in a very complex way
(\ref{3.13}), depending, in general, on the momenta of all the
particles that it meets in the vertex, while the interaction
coordinate $z^\alpha$ is becoming translated by a constant, just
like in special relativity.

It is considerably harder to prove translational invariance for a
`tree process' with many vertices (meaning the particles created at
one vertex to not interact again), and there are indications that
when loop processes are allowed the translational invariance might
be inevitably lost (see Ref.\ \cite{AmelinoCamelia:2011nt} for more
details.) In the multi-vertex situation, the translational
invariance is in a sense `holistic' since its action on a particular
particle wordline carries an information about all other particles
in the universe the particle in question has interacted or will
interact with.

Let us now turn to Lorentz symmetry. It should be noted that since,
as discussed in the preceding section, one uses different and a
priori independent geometric objects to construct free action
(\ref{3.7}) and the interaction term (\ref{3.11}) Lorentz invariance
is manifestly violated in general. As we will see in a moment it is
always possible to make the free action Lorentz invariant; then the
interaction term may turn out to be invariant as well, or not (for
example one may have a nontrivial metric,, and therefore a
nontrivial mass-shell relation and the standard linear conservation
law.) In what follows we will investigate under which condition the
total action is Lorentz invariant.

Let us assume for simplicity that the action of rotational subgroup
of the Lorentz group is un-deformed (which means that the momentum
space tetrad and the connection transform as tensors of appropriate
rank under the action of rotations). Consider the mass-shell
relation in the free action first. It is Lorentz invariant if there
exist three `generators' (Hamiltonian vector fields) $N_i$
satisfying the Poisson bracket relations
\begin{equation}\label{3.14}
    \{M_i, N_j\}=\epsilon_{ij}{}^k\, N_k\,,\quad \{N_i, N_j\}=-\epsilon_{ij}{}^k\,
    M_k\,,
\end{equation}
where $M_i$ are generators of rotations $\{M_i,
M_j\}=\epsilon_{ij}{}^k\, M_k$, and such that
\begin{equation}\label{3.15}
  \delta_\lambda {\cal C}(p)\equiv \lambda^i \{N_i, {\cal C}(p)\}=0\,.
\end{equation}
It is easy to check that such generators always exist. Notice now
that since ${\cal C}(p)$ is an integral of $g^{\alpha\beta}(p)\,
\dot p_\alpha\dot p_\beta$ with $g^{\alpha\beta}(p) = \eta^{ab}\,
E^\alpha_a(p)\, E^\beta_b(p)$, it follows from (\ref{3.15}) that
$E^\alpha_a(p)\,\dot p_\alpha$ must transform as a un-deformed
Lorentz vector, possibly with momentum dependent boost and rotation
transformations parameters (see Sect.\ 5 for an explicit example),
i.e.,
\begin{align}
  \delta_\lambda E^\alpha_0(p)\,\dot p_\alpha&=\lambda^i   \{N_i,E^\alpha_0(p)\,\dot p_\alpha\} =\bar\lambda^i(p)\, E^\alpha_i(p)\,\dot
    p_\alpha\,,\newline\\  \delta_\lambda E^\alpha_j(p)\,\dot p_\alpha&=\lambda^i \{N_i,E^\alpha_j(p)\,\dot p_\alpha\} =\bar\lambda_{j}(p)\, E^\alpha_0(p)\,\dot
    p_\alpha+ \epsilon_{ij}{}^{k}\,\bar\rho^j(p)\, E^\alpha_k(p)\,\dot p_\alpha\,.\label{3.16}
\end{align}
Then it follows that $x^a$ transform under momentum dependent
Lorentz transformations as a vector as well
\begin{equation}\label{3.16a}
\delta_\lambda x^0 =- x^i\bar\lambda_i(p)\,,\quad \delta_\lambda x^i
=- x^0\bar\lambda^i(p) +\epsilon{}^i{}_{jk} \bar\rho^j(p)\, x^k\,,
\end{equation}
  and the
free action is Lorentz-invariant. The transformations  (\ref{3.16a})
are just the standard Lorentz transformations but with momentum
dependent transformation parameter.

Having checked the invariance of the free action let us turn to the
interaction term. There are three possibilities.

It may happen that ${\cal K}_\alpha$ just transform covariantly
under the action of Lorentz symmetry defined above, i.e.,
$$\delta_\lambda {\cal K}_\alpha = \lambda^i \Lambda_i{}_\alpha^\beta\, {\cal K}_\beta\,,$$
where $\Lambda_i$ is a $4\times4$ Lorentz matrix representing the
boost. This is the case, for example in 2+1 gravity discussed in the
preceding section (in this case the on-shell relation and
$E^\alpha_a(p)\,\dot p_\alpha$ are manifestly Lorentz scalar and
vector, respectively.) Such possibility corresponds to the
un-deformed Lorentz symmetry with a nontrivially deformed
translational sector.

Another possibility is that ${\cal K}_\alpha$ cannot be made
covariant under the action of (deformed) Lorentz symmetry that
leaves invariant the free action. In this case Lorentz symmetry is
manifestly violated, the momentum conservation rule ${\cal
K}_\alpha=0$ holds only in one specific Lorentz frame. As another
side of the same coin one may consider the situation, in which
${\cal K}_\alpha$ is the standard linear combination of momenta
(\ref{3.5}) but the on-shell relation (and the form of the momentum
space tetrad) violates the (standard) Lorentz symmetry manifestly,
which is the standard formulation of the theories with Lorentz
invariant violation. The two situations described above in this
paragraph are related by a change of coordinates in momentum space.

The third possibility is the most interesting as it uses the
techniques of Hopf algebras.\cite{Majid:2006xn,Gubitosi:2011ej}
Consider first two particles. Their total momentum $P_\alpha$ is
given by a deformed sum of the particles' momenta
\begin{equation}\label{3.17}
    P_\alpha = (p\oplus q)_\alpha
\end{equation}
We demand that this expression is Lorentz-covariant, i.e., the left
hand side of eq.\ (\ref{3.17}) transforms exactly like the right
hand side
\begin{equation}\label{3.18}
    (\delta_\lambda P)_\alpha = \left[\delta_\lambda(p\oplus q)\right]_\alpha
\end{equation}
If we assume that $\delta_\lambda$ on the left hand side satisfies
the Leibniz rule usually this condition cannot be met because
$$
\left[\delta_\lambda(p\oplus q)\right]_\alpha\neq
\left[\delta_\lambda(p)\oplus q\right]_\alpha+\left[p\oplus
\delta_\lambda(q)\right]_\alpha\,.
$$
However it may happen that this equation could be satisfied if we
relax the Leibniz rule allowing the parameter $\lambda$ in the
second term to be replaced by momentum dependent one $\lambda
\triangleleft p$ (and allowing for the action of another symmetries
belonging to the full group in question, in our case the rotations)
so that\footnote{This can be understood by noticing that if the
relation $\oplus$ is nontrivial, it must depend on $p$ and $q$ in a
nontrivial way, and thus it is, in general, it cannot be inert under
Lorentz transformations.}
\begin{equation}\label{3.19}
\left[\delta_\lambda(p\oplus q)\right]_\alpha=
\left[\delta_\lambda(p)\oplus q\right]_\alpha+\left[p\oplus
\delta_{\lambda\triangleleft p}(q)+ \ldots\right]_\alpha\,,
\end{equation}
where $\ldots$ denote other terms that might be present (see Sect.\
5 for an explicit example.) Because of the close relation of the
Lorentz symmetry action to the co-product structure known from the
theory of Hopf algebras we say that the action (\ref{3.19}) is
characterized by a nontrivial coproduct. If (\ref{3.19}) is
satisfied the composition rule is covariant and both sides of
(\ref{3.17}) transform under Lorentz symmetry action in the same
way.

There are some consistency conditions that must be satisfied in
order for (\ref{3.19}) to hold. First, assuming associativity of the
composition rule\footnote{This construction is based on the theory
of Hopf algebras, in which associativity is assumed. To handle the
non-associative case one would presumably have to employ Hopf
algebroids.\cite{xu}} we have
$$
\delta_\lambda((p\oplus q)\oplus r) = \delta_\lambda(p\oplus
(q\oplus r))
$$
so that
\begin{equation}\label{3.20}
    \delta_{\lambda \triangleleft (p\oplus q)} = \delta_{(\lambda \triangleleft p)\triangleleft
    q}+\ldots\,.
\end{equation}
Moreover assuming Lorentz invariance of the identities $(\ominus
p)\oplus p=0 $, $p\oplus (\ominus p)=0 $ we have
\begin{equation}\label{3.21}
    \delta_\lambda(\ominus p) \oplus p + (\ominus p) \oplus\delta_{\lambda\triangleleft \ominus
    p} p=0=\delta_\lambda p \oplus(\ominus p) + p \oplus\delta_{\lambda\triangleleft
    p} (\ominus p)\,.
\end{equation}
We will discuss all this notions explicitly in Sect.\ 5 taking as na
example the case of $\kappa$-Poincar\'e based construction.

The last symmetry of the action we will briefly discuss is the
invariance with respect to the change of coordinates in momentum
space. The invariance of the free action is again straightforward to
see: both the on-shell relation and the components of the expression
$E^\alpha_a \, \dot p_\alpha$, $a=0,\ldots,3$ behave, by
construction, as scalars under diffeomorphisms.

It is a nontrivial problem, however, to find out how the momentum
conservation rule ${\cal K}_\alpha$ transforms under the change of
coordinates. To understand this let us reflect on the notion of
total momentum of two particles
$$
p^{tot}_\alpha (p^1, p^2)=( p^1\oplus p^2)_\alpha\,.
$$
The physical meaning of this expression is that there are to ways of
looking at the momentum of the composite system: one can see this
system as combination of two particles with momenta $p^1$ and $p^2$
or as a single point particle with momentum $p^{tot}$. Now, when we
change variables $p\mapsto P(p)$, where $P(p)$ is an invertible
function expressing the new variables in terms of the old ones, the
same duality between the total momentum and the momenta of
constituents must hold. This means that the total momentum in new
variables is the same function of $p^{tot}$ as $P$ is a function of
$p$. Therefore
\begin{equation}\label{3.22}
p^1\oplus p^2=p^{tot}\mapsto P^{tot}(p^{tot})
=P^{tot}\left(\left.p^1\oplus p^2\right|_{p^{1/2} =
p^{1/2}(P^{1/2})}\right)
\end{equation}
To see this rule in action consider the following change of
variables in 2 dimensions $(p_0,p) \mapsto (P_0,P)$ of the
form\footnote{This example illustrates the transition from the
bicrossproduct to the classical basis in $\kappa$-Poincar\'e; see
Sect.\ 5 for more details.}
$$
P_0 =  \kappa\,\sinh {{p_0}/\kappa} + \frac{p^2}{2\kappa}\, e^{
{p_0}/\kappa}\,, \quad
 P=   p \, e^{  {p_0}/\kappa}\,.
 $$
Let the composition rule in $p$ variables read
$$
(p^1\oplus p^2)_0 =p^1_0+ p^2_0\,,\quad  (p^1\oplus p^2) = p^1 +
e^{- {p^1_0}/\kappa}\, p^2\,.
$$
Therefore
\begin{eqnarray*}
 P^{tot} &=&\left(p^1(P^1) + e^{- {p^1_0}(P^1)/\kappa}\,
p^2(P^2)\right)e^{ ( {p^1_0}(P^1)+{p^2_0}(P^2) )/\kappa}   \nonumber\\
   &=& P^1 \,e^{ {p^2_0}/\kappa}(P^2) + P^2 =(P^1\oplus P^2)_1\,.\nonumber
   \end{eqnarray*}
Similar expression can be derived for the zeroth component of the
total momentum. It should be noticed that the above definition of
how the composition of momenta transforms under diffeomorphisms
agrees with the rule of transformation of coproduct of a Hopf
algebra under change of basis, as it should. It is worth noticing
that when one uses the expansion near the origin of the momentum
space $\cal O$
$$
(p^1\oplus p^2)_\alpha =p^1_\alpha+ p^2_\alpha +
\Gamma_\alpha{}^{\beta\gamma}\, p^1_\beta\, p^2_\gamma
$$
and makes a change of coordinates according to recipe described
above, one finds that $\Gamma_\alpha{}^{\beta\gamma}$ transforms as
a connection, which justifies dubbing
$\Gamma_\alpha{}^{\beta\gamma}$ the connection
coefficients\cite{AmelinoCamelia:2011bm,Freidel:2011mt}).

Similar definition holds for the operation $\ominus$ and using them
one can reconstruct the form of the conservation rule for an
arbitrary vertex after change of momentum space variables.

This concludes our discussion of the particle action and its
symmetries.

\section{Spacetime and relative locality}

In the previous sections we concentrated our investigations on
curved momentum spaces. Let us now investigate what is the fate of
spacetime in theories, in which the momentum space is nontrivial. As
we will see below  it turns out to be the major difference between
the spacetimes of special (and general) relativity  and the present
case is the necessity to replace the absolute notion of locality
valid for all observers  with the somehow relaxed notion of relative
locality.

It should be stressed at the very beginning that in order to
formulate the free action (\ref{3.4}) or (\ref{3.7}) we must have in
our disposal the spacetime structure, with coordinates $x^a$ defined
in such a way that the wordlines of free particles are straight
lines. This is essentially guaranteed in special relativity because,
operationally, to define coordinates one uses freely propagating
light signals\footnote{In principle one is free to use some other
freely moving probes, but for the sake of concreteness we will
discuss only light signals here.} and the standard coordinates are
adjusted so as to make the wordlines straight.

In the curved momentum space case (\ref{3.7}) the situation is more
complex, because the coordinatizations done with the help of light
signals of different energies will lead, in general, to different
resulting coordinates. One can circumvent this problems by using
light signals with an  appropriately low energy, so as to make the
impact of a nontrivial momentum space geometry negligible. In
classical theory, where we think of the light signal in terms of a
bunch of pointlike massless particles, this is, of course, not
problematic. In quantum theory, however, such procedure is
questionable because quantum mechanically we  have to do with waves,
not point particles, and therefore the resolution of the
low-energetic probes is bounded from below by their wavelength. This
causes a fundamental problem, but for all practical purposes it is
not really relevant, because if the scale $\kappa$ is of order of
Planck energy, one can achieve, for example, the atom size
resolution by using photons with $p_0/\kappa \sim 10^{-25}$, for
whose the effects resulting from curved momentum space are
completely negligible.

Having defined the spacetime coordinates with the help of low energy
light signals we turn now to the question of how this spacetime is
related to the ones corresponding to  particles carrying momenta of
the magnitude comparable to the scale $\kappa$. Geometrically
speaking these spacetimes are spaces cotangent  to the momentum
manifold at some appropriate point ${\cal P}$ with coordinates
$p_\alpha$. Such spaces are isomorphic as vector spaces but there
is, in principle, no canonical isomorphism relating the spacetimes
at different points ${\cal P}$ and ${\cal Q}$ if ${\cal P}\neq{\cal
Q}$. It should be noticed however that this same problem appears
already in special relativity and is solved there by assuming that
the flat momentum space geometry possesses translational Killing
vectors which make it possible to translate rigidly all the
spacetimes to one point, the origin ${\cal O}$ of the momentum
space, say.

On the other hand, in general relativity one solves the analogous
problem of comparing tangent spaces at the different spacetime
points by using the tetrad $e^a_\mu(x)$ that maps all these spaces
into one `ambient' Minkowski space: $T_xM \ni v^\mu(x) \mapsto
e^a_\mu(x)\, v^\mu(x) = v^a \in {Mink}_4$.

We made use of the same trick in constructing our free action
(\ref{3.7}). There was a momentum space tetrad there that
effectively made the coordinates $x^a$ belong to a single Minkowski
spacetime, independently of the point in the momentum space.
Moreover, since the curved momentum space effect are to vanish at
the momentum space origin ${\cal O}$ (by correspondence principle
which guarantees that at appropriately small momenta our theory
becomes identical with special relativity) this Minkowski spacetime
is just the ones we coordinatized with the help of the low energy
light signals, as described above. Having constructed the positions
all belonging to the same spacetime we are now free to compare
positions of wordlines of particles carrying different momenta.
Moreover, since as we have shown in Sect.\ 3, if the Lorentz
symmetry is present the coordinates $x^a$ transform as Lorentz
vectors, one can equip the spacetime with the standard Minkowski
metric, which makes it possible to compute distances between points
in the standard way.

There is a prize to be paid, however. The kinetic term in
(\ref{3.7}) is non-trivial, which leads to the `non-commutativity'
of spacetime coordinates, expressed in the classical theory by the
fact that their Poisson bracket is not equal to zero. This can be
seen directly by calculating the symplectic form and then the
Poisson brackets, but here we will use a simpler way to compute
them.

Instead of the $x^a$ variables,  being coordinates in the universal
spacetime we discussed above (\ref{3.7}), let us introduce the
cotangent space ones by
$$
x^\alpha(p) \equiv x^a\, E^\alpha_a(p)\,.
$$
In terms of $x^\alpha$ the kinetic term is $-x^\alpha\dot p_\alpha$
so that the only non-vanishing Poisson bracket reads
$$
\left\{ x^\alpha, p_\beta \right\} = \delta^\alpha_\beta\,.
$$
Returning to the physical coordinates using the inverse tetrad $x^a
= E^a_\alpha(p) x^\alpha$ we find
\begin{align}
\left\{ x^a, p_\beta \right\} &= E^a_\beta(p)\label{4.1}\\
\left\{ x^a, x^b \right\} &=  E^a_\alpha\,E^b_\beta\left(
 E^{\alpha,\beta}_c - E^{\beta,\alpha}_c\right)x^c\,,\label{4.2}
\end{align}
with momenta having  vanishing Poisson bracket, so that indeed in
the case of any non-trivial momentum space geometry the position
variables have a non-vanishing Poisson bracket. This is a nice
illustration of the idea that curved momentum spaces are in
one-to-one relation with non-commutative
geometry\cite{Majid:1999tc}.

Having discussed the spacetime aspect of the free action, let us now
turn to the interaction part. Here we encounter the notion of {\em
relative locality}. Let us take a vertex with some wordlines coming
in and some going out, which is characterized by the interaction
coordinates $z^a$. The equation following equation (\ref{3.12})
relates the coordinate of the end of the wordline of a particle
carrying momentum $p$ with the interaction one
\begin{equation}\label{4.3}
    x^a(0) = E^a_\beta(p)\, z^\alpha\, \frac{\partial{\cal
    K}_\alpha(p,p^{(1)},\ldots)}{\partial p_\beta}\,.
\end{equation}
The first thing which should be noticed is that if the momenta of
all the particles are very small compared to the scale $\kappa$,
${\cal     K}_\alpha(p,p^{(1)},\ldots) \sim
p_\alpha+p^{(1)}_\alpha+\ldots$, $E^a_\beta(p)\sim \delta^a_\beta$
and the wordline and interaction coordinates coincide. Moreover if
the interaction event takes place at the origin of a coordinate
system, of Alice, say, so that $z^\alpha=0$ it follows from
(\ref{4.3}) that $x^a(0)=0$ as well, for arbitrary interaction and
the magnitude of momenta.

This changes however when the interaction event takes place away
from the origin of the Alice's coordinates. Then $x^a(0) \neq
z^\alpha$. But the spacetime point with coordinates $z^\alpha$ is an
origin of some another, translated coordinate system, say of Bob,
and by the same equation (\ref{4.3}), for Bob $x^a(0) =z^\alpha=0$.
This shows that in the theories with curved momentum space the
concept of locality loses its absolute status and becomes a notion
relative to the observer.

This may sound as a very radical departure from the standard
physical intuitions, but it should be noted that operationally
speaking all the physical experiments are performed at the origin of
the observer's coordinate system, and the knowledge of distant
events is only inferred with the help of some (light or otherwise)
signals sent from the events to the observer. What really matters is
whether the relativity principle holds, i.e., whether the pictures
of a physical process inferred by different observers are mutually
consistent, so that neither of them can be treated as a privileged
one.

The relative feature of locality has been nicely illustrated in the
recent paper\cite{AmelinoCamelia:2011cv}, in the framework of
$\kappa$-Poincar\'e. Consider an observer Alice who sends two light
signals of different energies (one low-energetic and second of very
high energy) in the direction of a distant observer Bob, who is
static with respect to Alice. According to Alice these two signals
reach Bob at exactly the same spacetime point, because they were
sent simultaneously from her location and, in $\kappa$-Poincar\'e
the velocity of massless particles is energy-independent (see next
Section.) On the other hand Bob does not see these two particles
arriving simultaneously, for him there is a time lag between arrival
time of order of $D \, \Delta E/\kappa$, where $D$ is the distance
between Alice and Bob and $\Delta E$ the difference of particles'
energies. How this can be reconciled with the fact that for Bob too
all massless particles move with the universal speed of light? The
answer turns out to be simple: for Bob the events of particles'
creation at Alice's location are not simultaneous, but there is a
time lag between them, exactly the same as the time lag of the
particles' arrival time. Thus even if Alice does not agree with Bob
about which events are local and which are not, their points of view
are perfectly mutually consistent, in a relativistic way.

A similar analysis can be made in the case of quantum
mechanics\cite{AmelinoCamelia:2012ra}. Here we also have two distant
observers, Alice and Bob, at relative rest, who create two identical
Gaussian wave packets at their origins. In Alice's description the
packet at her origin is perfectly spherically symmetric, but the one
at Bob's has larger `fuzziness.' According to relative locality,
Bob's view should be exactly the same, just with wave packets
changing roles. The direct computation presented
in\cite{AmelinoCamelia:2012ra} confirms this expectation.

Let us finish this section asking a question if relative locality is
a real physical effect or perhaps just a (spacetime) coordinate
artifact? As noted above even special relativity can be artificially
made a theory with relative locality, by replacing the standard
Minkowski space coordinates with some new, momentum dependent ones
(although in order to do so one has to introduce a mass scale, which
is not really available in special relativity.) Then the question
arises if in the theories with curved momentum space the relative
locality property can be undone by changing position variables?
There is no general answer to this question; it can be checked
however that in the case of $\kappa$-Poincar\'e construction, to be
discussed in more details in the next section, when the nontrivial
structure of momentum space and deformation of spacetime symmetries
are both `genuine', i.e., cannot be undone by any change of
variables, relative locality is genuine too.

This concludes our brief presentation of the role of spacetime in
the theories with curved momentum space and the principle of
relative locality.

\section{$\kappa$-Poincar\'e}

Till now we we discussed the particles' model in a rather abstract
way, without referring to any particular example. Let us therefore
turn now to a specific model, which has its roots in
$\kappa$-Poincar\'e algebra and $\kappa$-Minkowski space
constructions\cite{Lukierski:1991pn,Lukierski:1992dt,Lukierski:1993wx,Majid:1994cy}.
In what follows we will borrow from the presentations in Refs.\
\cite{Freidel:2007hk} and \cite{Gubitosi:2011ej}.

Let us start with describing the momentum space of a
$\kappa$-Poincar\'e particle. It is a four dimensional group
manifold of a Lie group $AN(3)$, whose Lie algebra generators
satisfy
\begin{equation}\label{5.1}
 [X^0, X^i] = \frac{i}{\kappa}\,  X^i\,.
\end{equation}
Sometimes $X^a$ above are interpreted as positions in  a
non-commutative spacetime; such spacetime is known under the name of
`$\kappa$-Minkowski space'\cite{Lukierski:1993wx}. For our purposes
the relevant matrix representation of this Lie algebra will happen
to be the 5-dimensional one, in which case we have
\begin{equation}\label{5.2}
X^0 = -\frac{i}{\kappa} \,\left(\begin{array}{ccc}
  0 & \mathbf{0} & 1 \\
  \mathbf{0} & \mathbf{0} & \mathbf{0} \\
  1 & \mathbf{0} & 0
\end{array}\right) \quad
{\mathbf{X}} = \frac{i}{\kappa} \,\left(\begin{array}{ccc}
  0 & {\bm{\epsilon}\,{}^T} &  0\\
  \bm{\epsilon} & \mathbf{0} & \bm{\epsilon} \\
  0 & -\bm{\epsilon}\,{}^T & 0
\end{array}\right),
\end{equation}
where bold fonds are used to denote space components of a 4-vector
and $\bm{\epsilon}$ is a three dimensional vector with a single unit
entry, e.g., $\epsilon^1 = (1,0,0)$. This algebra has one abelian
generator $X^0$ and three nilpotent generators $X^i$ (one can check
that $(X^i)^3=0$) which is the reason behind the notation
$AN(3)$\footnote{In the $\kappa$-Poincar\'e literature this algebra
is usually called the $\kappa$-Minkowski algebra and is defined as a
dual (in the Hopf algebraic sense) to the $\kappa$-Poincar\'e Hopf
algebra, see Refs.\ \cite{Lukierski:1993wx,Majid:1994cy} and
\cite{Borowiec:2010yw} for more details and additional references.}.
Let us now consider a group element of $AN(3)$ (sometimes called the
`ordered plane wave on $\kappa$-Minkowski
space'\cite{AmelinoCamelia:1999pm})
\begin{equation}\label{5.3}
    \mathfrak{u}(p) =e^{ip_i  X^i} e^{ip_0  X^0}\,.
\end{equation}
In the representation (\ref{5.2}) this is a $5\times 5$ matrix which
acts on $5$-dimensional Minkowski space as a linear transformation.
One finds
$$
\exp(i p_0 X^0)= \left(\begin{array}{ccc}
  \cosh\frac{p_0}\kappa & \mathbf{0} & \sinh\frac{p_0}\kappa \\&&\\
  \mathbf{0} & \mathbf{1} & \mathbf{0} \\&&\\
  \sinh\frac{p_0}\kappa\; & \mathbf{0}\; & \cosh\frac{p_0}\kappa
\end{array}\right)\, , \quad
\exp(i p_i X^i) = \left(\begin{array}{ccc}
  1+\frac{\mathbf{p}^2}{2\kappa^2}\; &\; \frac{\mathbf p}\kappa \;& \; \frac{\mathbf{p}^2}{2\kappa^2}\\&&\\
  \frac{\mathbf p}\kappa & \mathbf{1} & \frac{\mathbf p}\kappa \\&&\\
  -\frac{\mathbf{p}^2}{2\kappa^2}\; & -\frac{\mathbf p}\kappa\; & 1-\frac{\mathbf{p}^2}{2\kappa^2}
\end{array}\right)\, ,
$$
where $\mathbf{1}$ is the unit $3\times 3$ matrix from which we find
\begin{equation}\label{5.3a}
 \mathfrak{u}(p)   =\left(\begin{array}{ccc}
 \frac{ \bar P_4}\kappa \;&\; e^{-p_0/\kappa}\, \frac{\mathbf P}\kappa \; &\;
\frac{  P_0}\kappa\\&&\\
  \frac{\mathbf P}\kappa  & \mathbf{1} & \frac{\mathbf P}\kappa  \\&&\\
  \frac{\bar P_0}\kappa\; & -e^{-p_0/\kappa}\, \frac{\mathbf P}\kappa\; &
\frac{  P_4}\kappa
\end{array}\right)\, .
\end{equation}

 We now take a special point in this space, the origin
$\cal O$ with coordinates $(0,\ldots,0,\kappa)$\footnote{Had we
chosen another point as ${\cal O}$, we would get different
realizations of the group $AN(3)$. For example if ${\cal O}=(\kappa
,\ldots , 0)$ we would get the  Euclidean realization and for ${\cal
O}=(\kappa,0,0,0,\kappa)$ the light-cone one.} and act with
(\ref{5.3a}) on it. As a result we get a point in the
$5$-dimensional Minkowski space with coordinates $(P_0, {P}_i, P_4)$
with
\begin{eqnarray}
 {P_0}(p_0, \mathbf{p}) &=&\kappa  \sinh
\frac{p_0}{\kappa} + \frac{\mathbf{p}^2}{2\kappa}\,
e^{  {p_0}/\kappa}, \nonumber\\
 P_i(p_0, \mathbf{p}) &=&   p_i \, e^{  {p_0}/\kappa}, \nonumber\\
 {P_4}(p_0, \mathbf{p}) &=& \kappa \cosh
\frac{p_0}{\kappa} - \frac{\mathbf{p}^2}{2\kappa}\, e^{
{p_0}/\kappa}. \label{5.4}
\end{eqnarray}
It is easy to check that
\begin{equation}\label{5.5}
   -P_0^2 + \mathbf{P}^2 + P_4^4 =\kappa^2
\end{equation}
and therefore the group $AN(3)$ is isomorphic, as a manifold to
$4$-dimensional de Sitter space. Actually, this group is not the
whole of this space, but rather a half of it because it follows from
(\ref{5.4}) that
\begin{equation}\label{5.6}
    P_0+P_4 = e^{p_0/\kappa}>0\,,\quad P_4\equiv \sqrt{\kappa^2 +P_0^2 +\mathbf{P}^2}>0\,.
\end{equation}

To construct the free action with the group $AN(3)$ as a momentum
space, let us recall that in this case there exists a canonical form
of the kinetic term provided by the Kirillov symplectic
form\cite{Kirillov}. It can be constructed as follows. Since the
positions belong to the cotangent space of the momentum space they
can be naturally associated with elements of a dual to the Lie
algebra, which is a linear space spanned by the basis $\sigma_a$
with the pairing
$$
\left<\sigma_a, X^b\right>=\frac1i\,\delta_a^b\,.
$$
Now having a group element $\mathfrak{u}(p)$ (\ref{5.3}) we know
that $\mathfrak{u}^{-1}\dot{ \mathfrak{u}}$ is an element of the Lie
algebra $AN(3)$ and thus we can define the kinetic term of the
action as\footnote{Had we chosen to use $\dot{
\mathfrak{u}}\mathfrak{u}^{-1}$ instead we would get an action
$S^{kin} = -\int     d\tau\, x^\alpha\dot{p}_\alpha -
(\kappa)^{-1}\, x^i\, p_i\, \dot{p}_0$, which is
    related to (\ref{5.7}) by the change of variables, $\mathbf{p}\mapsto
    \tilde{\mathbf{p}}=\mathbf{p}
    e^{p_0/\kappa}$.}
\begin{equation}\label{5.7}
    S^{kin} =-\int d\tau\, x^a\left<\sigma_a, \dot{ \mathfrak{u}}\mathfrak{u}^{-1} \right>= -\int
    d\tau\, x^0\dot{p}_0 + e^{p_0/\kappa}\, x^i\,  \dot{p}_i\,,
\end{equation}
where the last equality is obtained by substituting the explicit
form of $\mathfrak{u}(p)$, (\ref{5.3}). From (\ref{5.7}) we can
immediately read of the components of the momentum space tetrad
\begin{equation}\label{5.8}
    E_0^0 = 1\,,\quad E_i^j= e^{p_0/\kappa}\,\delta_i^j\,,
\end{equation}
and the  line element takes the form
\begin{equation}\label{5.9}
    ds^2 =-  dp_0^2  +e^{2p_0/\kappa}\,d\mathbf{p}^2\,,
\end{equation}
where  $d\mathbf{p}^2\equiv dp_idp_i$,  which is nothing but the
metric of de Sitter space in `flat' coordinates. It should be
noticed that the metric (\ref{5.9}) can be obtained as an induced
metric $ds^2= -dP_0^2 + d\mathbf{P}^2 + dP_4^2$ when (\ref{5.4}) is
used.

Having the metric we can calculate the distance function, which is
used to define the mass shell condition; it
reads\cite{Gubitosi:2011ej,AmelinoCamelia:2011nt}
$$
{\cal C}(p) = \kappa^2\,\mbox{arccosh } \frac{P_4}\kappa
$$
 so that the mass-shell condition reads
\begin{equation}\label{5.10}
     \cosh
\frac{p_0}{\kappa} - \frac{\mathbf{p}^2}{2\kappa^2}\, e^{
{p_0}/\kappa}= \cosh\frac{m}\kappa \,.
\end{equation}
Using (\ref{5.7}) and (\ref{5.10}) we construct the free particle
action
\begin{equation}\label{5.11}
    S_{free}^{\kappa P} = -\int d\tau\left( x^0\dot{p}_0 + e^{p_0/\kappa}\, x^i\,  \dot{p}_i +2 N\left[\kappa^2\cosh
\frac{p_0}{\kappa} - \frac{\mathbf{p}^2}{2}\, e^{ {p_0}/\kappa}-
\kappa^2\cosh\frac{m}\kappa\right]\right)\,,
\end{equation}
where we  replaced $N$ with $2N$ so that the action
$S_{free}^{\kappa P}$ reduces to the standard relativistic particle
action in the limit $\kappa\rightarrow\infty$. It is worth noticing
that the action (\ref{5.11}) leads to a nontrivial Poisson brackets
algebra
\begin{align}
    \left\{x^0, x^i\right\}&= -\frac1\kappa\, x^i\,, \quad\left\{x^i,
    x^j\right\}=0\nonumber\\
 \left\{x^0, p_0\right\}&= 1\,, \quad\left\{x^i,
    p_j\right\}= e^{-p_0/\kappa}\, \delta^i_j\,, \quad\left\{x^0,
    p_i\right\}=\left\{x^i,
    p_0\right\}=0\,,\label{5.12}
\end{align}
with $p_\alpha$ having vanishing Poisson brackets.

The equations of motion following from (\ref{5.11}) have the form
\begin{align}
\dot p_\alpha &=0\nonumber\\
\dot x^i &=-2N\, p_i\,,\nonumber\\
   \dot x^0 &=2N\left(\kappa\sinh\frac{p_0}{\kappa} -\frac{\mathbf{p}^2}{2\kappa}\, e^{
{p_0}/\kappa}\right)\,,\label{5.13}
\end{align}
supplemented by the mass shell condition (\ref{5.10}), where in the
equations for $x^a$ we omitted terms proportional to $\tau$
derivatives of momenta. It is worth noticing that the velocity of
massless particles
$$
\mathbf{v}^2=\left|\frac{\dot x^i}{\dot x^0}\right|^2=1
$$
so that the velocity of light is independent of the energy of
photons\cite{Daszkiewicz:2003yr}.
\newline

Having discussed the free part of the $\kappa$-Poincar\'e particle
action let us now turn to the interactions. In order to construct
the interaction term in the action, we must define how the momenta
are composed and what is the form of the inverse (antipode). They
can be directly inferred from the group composition and group
inverse, respectively.

Having two group elements $u(p)$ and $u(q)$ (\ref{5.3}) we can
multiply them forming another group element
\begin{equation}\label{5.14}
    u(p)u(q)\equiv u(p\oplus q)
\end{equation}
with
\begin{equation}\label{5.15}
    (p\oplus q)_0= p_0+q_0\,,\quad (p\oplus q)_i=
    p_i+e^{-p_0/\kappa}\,q_i\,.
\end{equation}
Similarly we define the antipode
\begin{equation}\label{5.16}
    u^{-1}(p)\equiv u(\ominus p)
\end{equation}
so that
\begin{equation}\label{5.17}
    (\ominus p)_0= -p_0\,,\quad (\ominus p)_i=
    -e^{p_0/\kappa}\,p_i\,.
\end{equation}
Clearly, $p\oplus(\ominus p) = (\ominus p)\oplus p=0$. Using these
two operations  and remembering that $\oplus$ is associative, we can
construct ${\cal K}_\alpha$ for an arbitrary vertex.

Thus the $\kappa$-Poincar\'e particle action has the form
\begin{equation}\label{5.18}
    S= \sum_{\cal I} S_{free}^{\kappa P} + \sum_{\cal V} z^\alpha {\cal
    K}_\alpha\,,
\end{equation}
where in the first term we have a sum of free $\kappa$-Poincar\'e
particle actions (\ref{5.11}) with an appropriate choice of the
integration ranges, and the second term is a sum of vertices
constructed with the help of $\oplus$ and $\ominus$ operations
(\ref{5.15}), (\ref{5.17}). This concludes the construction of the
$\kappa$-Poincar\'e particles' dynamics. Let us now turn to the
symmetries of the $\kappa$-Poincar\'e particles' system.

\subsection{$\kappa$-Poincar\'e symmetry}

The free $\kappa$-Poincar\'e particle actions (\ref{5.11}) is
invariant under both translational and Lorentz symmetries.

The explicit form of translational transformations that leave the
free action invariant in the absence of interactions can be easily
deduced from the general discussion presented in Sect.\ 3.1 and
reads
\begin{equation}\label{5.19}
    \delta x^0 = \xi^0\,,\quad \delta x^i = \xi^i\,
    e^{-p_0/\kappa}\,.
\end{equation}
It follows that the conserved Noether charges associated with the
translational symmetry are the components of the momentum, as usual.

In the case of a single vertex, the transformation rule  can be read
off from the general formula (\ref{3.13}) and read
\begin{equation}\label{5.20}
    \delta x^0_{\cal I}=\xi^\gamma\, \frac{\partial {\cal
    K}_\gamma}{\partial p_0^{\cal I}}\,,\quad
    \delta x^i_{\cal I}=\xi^\gamma\, \frac{\partial {\cal
    K}_\gamma}{\partial p_i^{\cal I}}\,e^{-p_0/\kappa}\,,\quad
    \delta z^\alpha = \xi^\alpha\,.
\end{equation}
The case of more than one vertex is much more complicated and is
discussed in details in\cite{AmelinoCamelia:2011nt}.

Let us now turn to Lorentz symmetry. The Lorentz transformations of
momenta $p_\alpha$ can be easily found if one recalls that the
components $P_\alpha$ in (\ref{5.4}) transform as components of a
Lorentz vector, while $P_4$ is a Lorentz scalar. This follows from
the fact that the four dimensional Lorentz group acts on four
dimension subspace of the five dimensional Minkowski space, in which
the $AN(3)$ group manifold is imbedded  in the standard linear way
and leaves the fifth dimension invariant. Thus we have
$$
\delta_\lambda P_0 = \lambda^i\, P_i\,,\quad \delta_\lambda P_i =
\lambda_i\, P_0\,, \quad\delta_\lambda P_4 =0\,.
$$
From these equations we deduce that $P_4$ is proportional to the
Casimir used in the mass-shell relation, which we indeed derived
above (\ref{5.10}). Further, using the chain rule we can calculate
from them the transformation laws of $p_\alpha$, which turn out to
be
\begin{equation}\label{5.21}
\delta_\lambda p_0 =\lambda^i p_i\,,\quad \delta_\lambda p_i
=\lambda_i
\left(\frac{\kappa}2(1-e^{-2p_0/\kappa})+\frac{\mathbf{p}^2}{2\kappa}\right)
- \frac1\kappa\, \lambda^j\,p_j\,p_i\,.
\end{equation}
Let us return to the line element (\ref{5.9}). One can check by
direct calculation that it is Lorentz invariant\footnote{This
follows immediately from the manifest Lorentz invariance of
$ds^2=-dP_0^2 + d\mathbf{P}^2 + dP_4^2$.}
\begin{equation}\label{5.22}
    \delta_\lambda\, ds^2 =0\,,
\end{equation}
and it can be further shown that
$$
 \delta_\lambda\, E^\alpha_0 \, dp_\alpha = \bar \lambda^i(p) \, E^\alpha_i \,
 dp_\alpha\,,\quad \delta_\lambda\, E^\alpha_i \, dp_\alpha = \bar \lambda_i(p) \,
 E^\alpha_0\, dp_\alpha + \epsilon_{ij}{}^k\, \bar\rho^j(p)\, E^\alpha_k \,
 dp_\alpha\,,
 $$
 with
 $$
 \bar\lambda_i(p) =e^{-p_0/\kappa}\, \lambda_i\,,\quad \bar\rho^j(p)
 = -\frac1\kappa\, \epsilon^j{}_{kl}\, \lambda^k\, p^l\,,
 $$
 from which we deduce that the positions transform in the way
 described in Sect.\ 3
 \begin{equation}\label{5.23}
 \delta_\lambda\, x^0  = -\bar\lambda_i(p) \, x^i \,,\quad \delta_\lambda\, x^i  = -\bar\lambda^i(p) x^0+\epsilon^i{}_{lm}\,\bar\rho^l\, x^m
 \,.
 \end{equation}

It can be checked by direct calculations that the transformations
(\ref{5.21}) and (\ref{5.23}) agree with the hamiltonian action of
the generators of boost $N_i$ on phase space
$$
\delta_\lambda (\star) = \lambda^i\left\{ N_i, \star\right\}\,,\quad
N_i = p_i x^0 +\left(\kappa\sinh\frac{p_0}\kappa
+\frac{\mathbf{p}^2}{2\kappa}e^{p_0/\kappa}\right)x_i -
\frac1\kappa\, p_i p_kx^k\, e^{p_0/\kappa}\,,
$$
with the Poisson bracket given by (\ref{5.12}).

 Having discussed the action of
Lorentz transformations in the case of the single particle free
action, let us now turn to the interaction term. We will discuss
only the composition law for two particles and the antipode; these
results can be easily generalized to the case of many particles.

Consider the composition law first. In order to secure the Lorentz
invariance it must be covariant i.e., in the equality
$$
p^{tot}_\alpha = (p\oplus q)_\alpha
$$
both sides must transform in exactly the same way. As we will see
this condition cannot be met if one insists on the Leibnizian action
of the symmetry transformations on the right hand side. Let us
recall the component form the composition law (\ref{5.15})
$$
p^{tot}_0 = p_0+q_0\,,\quad p^{tot}_i = p_i+e^{-p_0/\kappa}\, q_i
$$
It is clear that both sides of these equations are covariant with
respect to rotations: both sides of the equations are rotation
scalars in the first and components of a three-vector in the second.

Consider first the equation for the zeroth component of momentum.
The left hand side transforms as
$$
\delta_\lambda p^{tot}_0 = \lambda^i(p_i+e^{-p_0/\kappa}\, q_i)
$$
Following the general idea presented in Sect.\ 3 for the right hand
side we take
$$
\delta_\lambda(p_0+q_0) = \delta_\lambda p_0 +
\delta_{\lambda\triangleleft p} q_0
$$
 and we deduce that
$$ \delta_{ \lambda\triangleleft p} =
\delta_{\bar\lambda} + \ldots\,,\quad \bar\lambda_i
=e^{-p_0/\kappa}\,\lambda_i
 $$
where $\ldots$ denote possible terms that vanish when acting on
$q_0$. Such terms might be only proportional to rotations
$\delta_\rho$. Checking the covariance of the space part of the
composition law one winds that
\begin{equation}\label{5.24}
\delta_{ \lambda\triangleleft p} = \delta_{\bar\lambda} +
\delta_{\rho}\,,\quad \rho_i =\epsilon_i{}^{jk}\lambda_j\, p_k\,.
 \end{equation}

It remains to be checked what are the covariance properties of the
antipode $\ominus$. The covariance means that taking Lorentz
transformation first and the antipode of the result must be equal to
the action of a modified Lorentz transformation with the parameter
$\ominus\lambda$ on the antipode, i.e.,
$$
\delta_{\ominus\lambda} \ominus p =\ominus (\delta_\lambda p)\,.
$$
 Since the antipode
(\ref{5.17}) is manifestly rotationally covariant, we can just
concentrate on Lorentz boosts, as before. Then
$$
\delta_{\ominus\lambda} (\ominus p)_0 = -(\ominus\lambda)^i\, p_i =
\ominus (\lambda^i\, p_i)=-e^{p_0/\kappa}\,\lambda^i\, p_i\,,
$$
from which it follows that
$$
\ominus\lambda_i = e^{p_0/\kappa}\, \lambda_i + \ldots\,,
$$
where $\ldots$ denotes, as before, a possible contribution from
rotation. From the covariance of the spacial components of the
antipode one can identify this rotational transformation and the
final result is
\begin{equation}\label{5.25}
\delta_{\ominus\lambda} = e^{p_0/\kappa}\left(\delta_\lambda -
\delta_\rho\right)\,,\quad \rho_i =\epsilon_i{}^{jk}\, \lambda_j\,
p_k\,.
\end{equation}
It should be noticed that the derived action of Lorentz
transformations on the composition of momenta and the antipode is in
one-to-one correspondence with the co-product and the antipode of
Lorentz generators of the $\kappa$-Poincar\'e Hopf
algebra\cite{Lukierski:1991pn,Lukierski:1992dt,Lukierski:1993wx,Majid:1994cy}.

Having discussed the infinitesimal action of Lorentz transformations
let us finally turn to the apparent problem that seemingly plagues
the theory when we allow for the finite Lorentz transformations. It
was noticed already in the early paper\cite{Bruno:2001mw} and
discussed further in\cite{Majid:2006xn,Gubitosi:2011ej}. The problem
is that although the Lorentz transformations are well defined in the
case of positive anergy $p_0>0$ for all real values of the boost
parameter, for negative energy $p_0>0$ this is not the case and only
some finite interval of boost parameters is allowed. An easy way to
see this is to realize that the condition (\ref{5.6}) defining the
momentum space is clearly {\em not} Lorentz invariant. It is worth
stressing here that this problem is not really important in the
classical theory of particles, where the energies are always
positive. In the case of a field theory with $AN(3)$ momentum space
the problem seems to be indeed severe, and, if there, it would mean
that the  Lorentz invariant field theory with this kind of a
momentum space may not exist at all (see Refs.\
\cite{Freidel:2007hk} and \cite{Freidel:2007yu} for details.)

The solution of this apparent problem has been proposed in Ref.\
\cite{Arzano:2009ci}. The idea is to modify the way the Lorentz
transformation act on positive and negative energies with the
antipode action (with the parameter $\ominus\lambda$) employed in
the latter case. It turns out that such action is fully consistent
and free of problems mentioned above.

\section{Conclusions}

In this paper we reviewed motivations leading to and some properties
of the relativistic theory of particles with curved momentum space.
Let us conclude this presentation with the list of of the most
pressing problems that need to be solved before this theory reaches
a mature stage.

Although it is very encouraging that models with curved momentum
space can be rigorously derived from the theory of gravity coupled
to point particles in 2+1 spacetime dimensions one has to be able to
derive models with curved momentum space also in the physical 3+1
spacetime dimensions. As discussed in Sect.\ 2 there are indications
that such model do exist, but their explicit construction will
certainly put the whole framework on a much firmer ground.

In absence of such explicit construction it is quite important to
examine the internal consistency of models with curved momentum
space and their symmetries. As we mentioned in the main text already
in the case of tree interactions with many vertices it is quite hard
to prove the invariance of the model under global spacetime
translations. When the loop processes\footnote{As an example of a
loop process consider  some number of particles interacting in
vertex $A$ in such a way that two, or more, of the particles
interacting at $A$ interact again in another vertex $B$.} are
allowed, the preliminary investigations suggest that the spacetime
translational symmetry is lost, and, even worse, the very notion of
causality might be at stake\cite{Chen:2012fu}. It is not clear how
these results, if confirmed, are to be interpreted.

Also would be very interesting to investigate properties of
classical and especially quantum field theories in the case of
curved momentum space. The case of momentum space being a group
manifold is especially interesting, because in this case the
symmetries have a mathematical structure of a Hopf algebra, being an
appropriate deformation of the Poincar\'e algebra. If a consistent
interacting quantum field theory can be constructed, either in 3+1
or in 2+1 dimensions, it would serve as an explicit counterexample
of the celebrated Coleman-Mandula theorem.

Last but not least, as discussed above,  the theories with curved
momentum space enjoy the relative locality properties, which leads
to the prediction that the time of flight of light coming from
distant sources may become energy-dependent, with the time lag
proportional to the energy difference and distance. As we discussed
it is not clear that for photons the proportionality factor should
be of order of the inverse Planck energy, but if it is such effects
might be within reach of current
experiments\cite{AmelinoCamelia:2009pg,Rideout:2012jb}.

\section*{Acknowledgements}

I would like to thank M.\ Arzano and T.\ Trze\'sniewski for many
discussions and help with some calculations presented here. Thanks
are due also to A.\ Borowiec and especially  J.\ Lukierski for
asking right questions that let me improve the presentation greatly.

I thank G.\ Rosati for pointing out a mistake in the discussion of
Lorentz transformations in earlier version of this paper.

This work was supported in parts by the grant 2011/01/B/ST2/03354
and by funds provided by the National Science Center under the
agreement DEC- 2011/02/A/ST2/00294.

\end{document}